\documentclass[12pt]{article}

\usepackage{graphicx}

\begin {document}

\begin{center}

{\large \textbf{Masses and Magnetic moments of Triply Heavy Flavour Baryons in Hypercentral Model}} \\
\textbf{Bhavin Patel, Ajay Majethiya  and P. C. Vinodkumar}\\
Department of Physics, Sardar Patel University,
Vallabh Vidyanagar- 388 120, Gujarat, INDIA
\\
\end{center}

 \abstract{ Triply heavy
flavour baryons are studied using the hyper central description of
the three-body system. The confinement potential is assumed as hyper
central coulomb plus power potential with power index $p$. The
ground state ($J^P=\frac{1}{2}^+$ and $\frac{3}{2}^+$) masses of
heavy flavour
 baryons are computed for
different power index, $ p$  starting from 0.5 to 2.0. The predicted
masses are found to attain a saturated value
 with respect to variation in $p$ beyond the power index $p>1.0$.
 Using the spin-flavour structure of the constituting quarks and
 by defining effective mass of the confined quarks within
 the baryons, the magnetic moments are computed with no additional free
 parameters.}

\section{Introduction}
The investigation of properties of hadrons containing heavy quarks
is not only of great interest in understanding the dynamics of QCD
at the hadronic scale but also an interesting due to the rapid
progress with the discovery of hadronic resonances
 at different experimental groups like BaBar, CLEO, SELEX and
other B factories world over. The last generation of baryons within
the standard model are the triply heavy baryons and they are
heaviest composite states, predicted by the constituent quark model.
Essentially they are the $\Omega_{ccc}$, $\Omega_{ccb}$,
$\Omega_{cbb}$ and $\Omega_{bbb}$ baryons \cite{Gomshi2006}. After
the observation of the doubly charmed baryon by the SELEX
group\cite{Mattson2002}, it is expected that the triply heavy
flavour baryonic state may be in the offing very soon. The vital
properties of these heaviest baryons in Nature is their masses and
magnetic moments. This has generated much interest in the
theoretical predictions of their
properties\cite{Martynenko2007,Bjorken85,Jia2006,Silvestre1996,Migura2006,Amand2006}.
\\
Though considerable amount of data on the  properties  of the
singly-heavy baryons are available in literature
\cite{Silvestre1996,Bhavin2008,Mathur2002}, only sparse attention
has been paid to the spectroscopy of double and triple-heavy flavour
baryons, perhaps
mainly due to the lack of experimental incentives \cite{Jia2006}.\\
Theoretically, baryons are not only the interesting systems to study
the quark dynamics and their properties but are also interesting
from the point of view of simple systems to study three body
problems. Out of many approaches and methods available for the three
body
systems\cite{Mathur2002,Shifman1979,Wang2003,Godfrey1985,Kiselev1995,Richard1992,Tong2000,Roncaglia1995,Savage1990,Korner1994,Murthy1985,Roberts2007},
we employ here the hyper central approach due to its simplicity to
study the triply heavy flavour baryons. As in our earlier study of
single and double heavy flavour baryons
\cite{Bhavin2008,BhavinPramana2008}, for the present study of triply
heavy flavour baryons the confinement potential is assumed in the
hyper central co-ordinates of the coulomb plus power potential form.
The spin hyperfine interactions similar to the one employed
\cite{Bhavin2008,Garcilazo2007} but with the radial part expressed
in terms of the hyper co-ordinate $x$ has been used with explicit
mass dependence of the constituting quarks in the present study. The
magnetic moments of heavy flavour baryons are computed based on the
nonrelativistic quark model using the spin-flavour wave functions of
the constituting quarks. The binding energy effects are considered
by defining an effective mass of the bound quarks within the baryon
for computing the magnetic moments. In the present study, we compute
the effective masses of the constituent quarks within the baryon
with different combinations of heavy flavour quarks. We repeat our
computations by varying the confinement potential power index $p$
from 0.5 to 2.0 to have an idea about the most suitable form of
interquark potential that yields the static properties of the triple
heavy baryons.

\section{Hyper Central Scheme for Baryons}

Quark model description of baryons is a simple three body system of
interest. Generally the phenomenological interactions among the
three quarks are studied using the two-body quark potentials such as
the Isgur Karl Model\cite{Isgur1978}, the Capstick and Isgur
relativistic model\cite{Godfrey1985,Capstick1986}, the Chiral quark
model \cite{Harleen2003}, the Harmonic Oscillator
model\cite{Murthy1985,Roberts2007} etc. The three-body effects are
incorporated in such models through two-body and three-body
spin-orbit terms\cite{Bhavin2008,Garcilazo2007}. The Jacobi
Co-ordinates to describe baryon as a bound state of three
constituent quarks is given by \cite{Simonov1966}
\begin{equation}\label{}
    \vec{\rho}=\frac{1}{\sqrt{2}}(\vec{r}_1-\vec{r}_2)\,\,{;}\,\, \vec{\lambda}=\frac{1}{\sqrt{6}}(\vec{r}_1+\vec{r}_2-2\vec{r}_3)\\
\end{equation}
Such that
\begin{equation}\label{}
m_\rho=\frac{2\,\,m_1\,\, m_2}{m_1+m_2}
 \,\,{;}\,\,
m_\lambda=\frac{3\,\,m_3\,\,(m_1+m_2)}{2\,(m_1+m_2+m_3)}
\end{equation}
Here $m_1$, $m_2$ and $m_3$ are the constituent quark mass
parameters.\\ In the hypercentral model, we introduce the hyper
spherical coordinates which are given by the angles
\begin{equation}\label{}
\Omega_\rho=(\theta_\rho,\phi_\rho)\,\,{;}\,\,
\Omega_\lambda=(\theta_\lambda,\phi_\lambda)
\end{equation}\label{}
together with the hyper radius, $x$ and hyper angle $\xi$ respectively as,\\
\begin{equation}\label{}
x=\sqrt{\rho^2+\lambda^2}\,\,{;}\,\,\xi=\arctan\left(\frac{\rho}{\lambda}\right)\\
\end{equation}\label{}
the model Hamiltonian for baryons can be written as
\begin{equation}\label{eq:404}
H=\frac{P^2_\rho}{2\,m_\rho}+\frac{P^2_\lambda}{2\,m_\lambda}+V(\rho,\lambda)=\frac{P^2_x}{2\,m}+V(x)\\
\end{equation}\label{}
Here the potential $V(x)$ is not purely a two body interaction but
it contains three-body effects also. The three body effects are
desirable in the study of hadrons since the non-abelian nature of
QCD leads to gluon-gluon couplings which produce three-body forces
\cite{Santopinto1998}. Using hyperspherical coordinates, the kinetic
energy operator $\frac{P^2_x}{2\,m}$ of the three-body system can be
written as
\begin{equation}\label{}
\frac{P^2_x}{2\,m}=\frac{-1}{2\,m}\left(\frac{\partial^2}{\partial\,x^2}+\frac{5}{x}\frac{\partial}{\partial\,x}-\frac{L^2(\Omega_\rho,\Omega_\lambda,\xi)}{x^2}\right)\\
\end{equation}\label{}
where $L^2(\Omega_\rho,\Omega_\lambda,\xi)$ is the quadratic Casimir
operator of the six dimensional rotational group $O(6)$, and its
eigen functions are the hyperspherical harmonics, $Y_{[\gamma]l_\rho
l_\lambda}(\Omega_\rho,\Omega_\lambda,\xi)$ satisfying the
eigenvalue relation
\begin{equation}\label{}
L^2 Y_{[\gamma]l_\rho l_\lambda}(\Omega_\rho,\Omega_\lambda,\xi)=
\gamma(\gamma+4)Y_{[\gamma]l_\rho
l_\lambda}(\Omega_\rho,\Omega_\lambda,\xi)
\end{equation}\label{}
Here $\gamma$ is the grand angular quantum number and it is given by
$\gamma=2\nu+l_\rho+l_\lambda$, and $\nu=0,1,...$ and $l_\rho$ and
$l_\lambda$ being the angular momenta associated with the $\rho$ and
$\lambda$ variables.
\begin{table}[t]
\begin{center}\caption{Triply heavy baryon masses (masses are in
Me$V$)}\label{tab:01}

\begin{tabular}{clllll}
\hline
 Baryon & Model
&{$\textbf{J}^P=\frac{1}{2}^+$}&Others&{$\textbf{J}^P=\frac{3}{2}^+$}&Others\\
\hline

$\Omega^{++}_{ccc}$&$p=0.5$&$-$&$-$&4897&4803\cite{Martynenko2007}\\
&\ \ \ \ \ \ 0.7&$-$&$-$&4777&4790 {\cite{Bjorken85}} \\
&\ \ \ \ \ \ 1.0&$-$&$-$&4736&4760 {\cite{Jia2006}} \\
&\ \ \ \ \ \ 1.5&$-$&$-$&4728&4773 {\cite{Migura2006}} \\
&\ \ \ \ \ \ 2.0&$-$&$-$&4728&4777 {\cite{Bernotas2008}} \\
&&&&&4965 {\cite{Roberts2007}}\\
\\
 $\Omega^{+}_{ccb} $&$p=0.5$&8262&8018 \cite{Martynenko2007}& 8273&8025 \cite{Martynenko2007}\\
&\ \ \ \ \ \ 0.7&8132 &$-$& 8142&8200 {\cite{Bjorken85}}\\
&\ \ \ \ \ \ 1.0&8089&$-$&8099&7980 {\cite{Jia2006}}\\
&\ \ \ \ \ \ 1.5&8082&7984 {\cite{Bernotas2008}}&8092&8005 {\cite{Bernotas2008}}\\
&\ \ \ \ \ \ 2.0&8082&8245{\cite{Roberts2007}}&8092&8265{\cite{Roberts2007}}\\
\\
 $\Omega^{0}_{bbc} $&
$p=0.5$&11546&11280 \cite{Martynenko2007}&11589&11287\cite{Martynenko2007}\\
&\ \ \ \ \ \ 0.7&11400&& 11440&11480 {\cite{Bjorken85}} \\
&\ \ \ \ \ \ 1.0& 11354&& 11394&11190 {\cite{Jia2006}}\\
&\ \ \ \ \ \ 1.5& 11347&11139 {\cite{Bernotas2008}}&11386 &11163 {\cite{Bernotas2008}}\\
&\ \ \ \ \ \ 2.0& 11347&11535 {\cite{Roberts2007}}& 11386&11554 {\cite{Roberts2007}}\\
\\
$\Omega^{-}_{bbb} $&$p=0.5$&$-$&$-$&14688&14569\cite{Martynenko2007}\\
&\ \ \ \ \ \ 0.7&$-$&$-$&14504&14760 {\cite{Bjorken85}}\\
&\ \ \ \ \ \ 1.0&$-$&$-$&14451&14370 {\cite{Jia2006}}\\
&\ \ \ \ \ \ 1.5&$-$&$-$&14444&14276 {\cite{Bernotas2008}}\\
&\ \ \ \ \ \ 2.0&$-$&$-$&14444&14834 {\cite{Roberts2007}}\\
\hline\hline
\end{tabular}
\end{center}
\end{table}

\begin{figure}

\includegraphics[height=3.0in,width=2.9in]{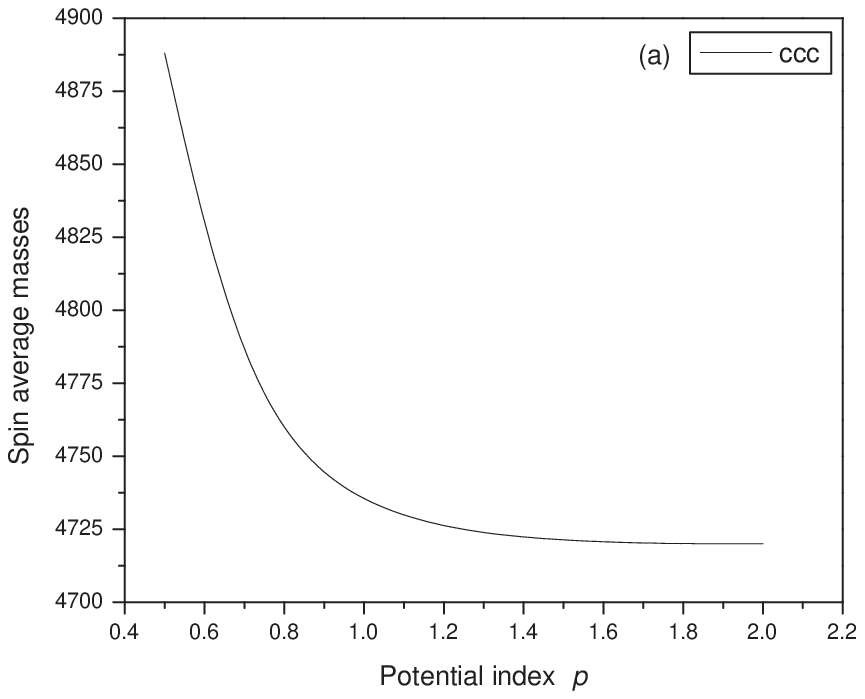}
\includegraphics[height=3.0in,width=2.9in]{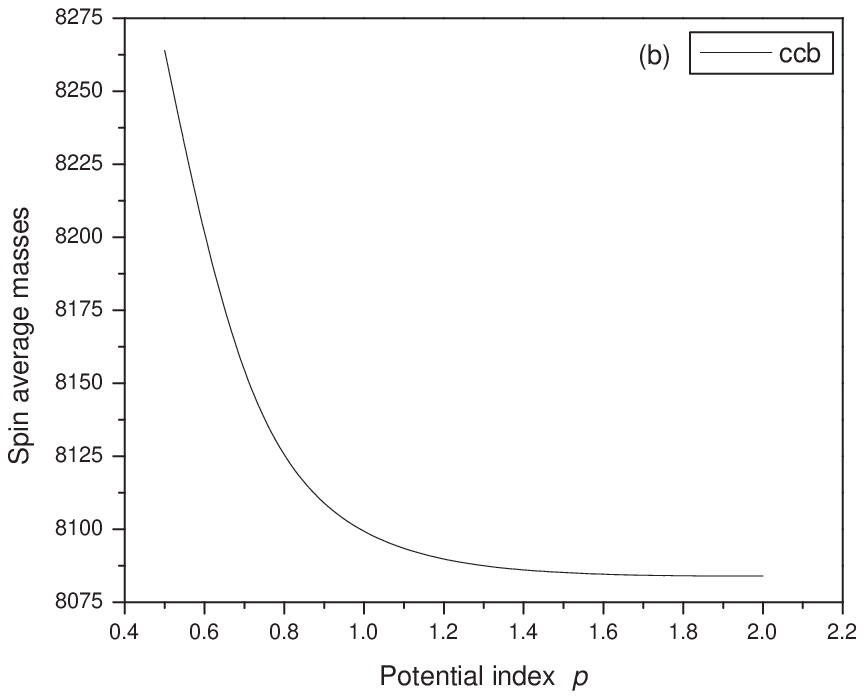}
\vspace{-0.2in} \vspace{-0.2in}
\includegraphics [height=3.0in,width=2.9in]{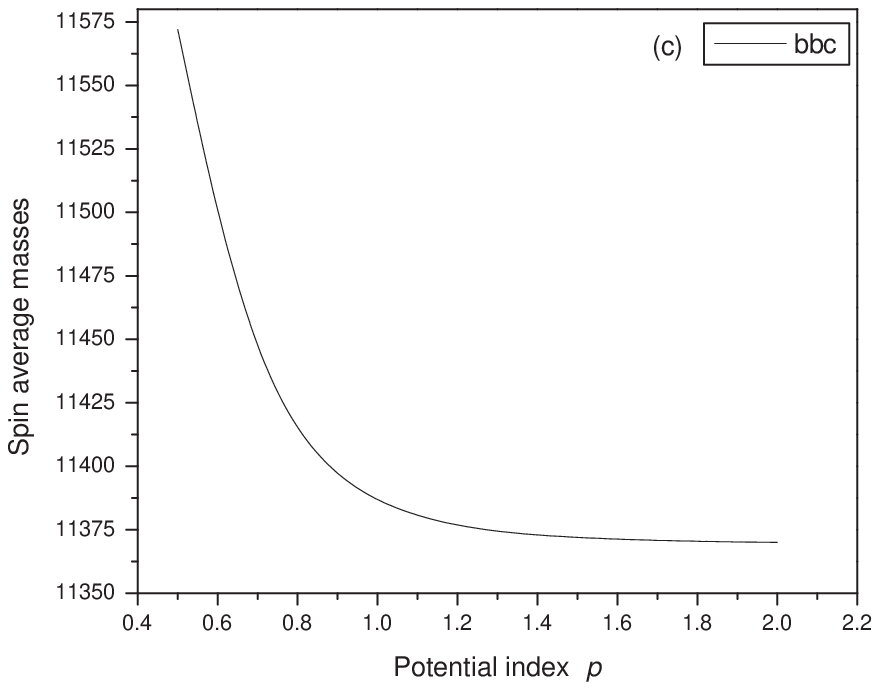}
\includegraphics [height=3.0in,width=2.9in]{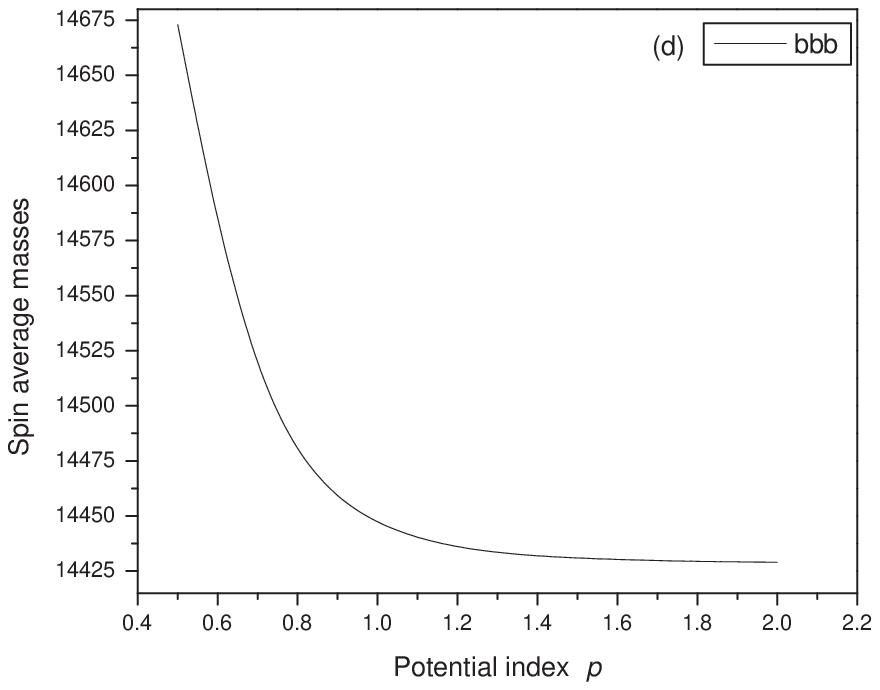}

\caption{\textbf{Variation of spin average masses (Me$V$) with
potential index $p$ for triply heavy baryons [a] $ccc$, [b] $ccb$,
[c] $bbc$ and [d] $bbb$.}}\label{fig:1}

\end{figure}

\begin{table}[t]
\caption{Spin-flavour wave functions and magnetic moments of heavy
flavour baryons with $J^{P}=\frac{1}{2}^{+}$} \label{tab:02}
\begin{tabular}{lllllll}
\hline
Baryon &Spin-flavour wave function&Megnetic moment\\
\hline
$\Omega^{*++}_{ccc}$&$c_{+}c_{+}c_{+}$&$3 \mu_{c}$\\

 $\Omega^{+}_{ccb}$&$\frac{\sqrt{2}}{6}(2b_{-}c_{+}c_{+}-
                                           c_{-}b_{+}c_{+}-
                                           b_{+}c_{-}c_{+}+
                                           2c_{+}b_{-}c_{+}-
                                           c_{+}c_{-}b_{+}$\\

 & \ \ \ \                                $-c_{-}c_{+}b_{+}-
                                           c_{+}b_{+}c_{-}-
                                          b_{+}c_{+}c_{-}+
                                           2c_{+}c_{+}b_{-})$&$\frac{4}{3}\mu_{c}-\frac{1}{3}\mu_{b}$\\
$\Omega^{*+}_{ccb}$&$\frac{1}{\sqrt{3}}(c_{+}c_{+}b_{+}+c_{+}b_{+}c_{+}+b_{+}c_{+}c_{+})$&$2\mu_{c}+\mu_{b}$\\

$\Omega^{0}_{bbc}$&$\frac{\sqrt{2}}{6}(2c_{-}b_{+}b_{+}-
                                           b_{-}c_{+}b_{+}-
                                           c_{+}b_{-}b_{+}+
                                           2b_{+}c_{-}b_{+}-
                                           b_{+}b_{-}c_{+}$\\

 & \ \ \ \                                $-b_{-}b_{+}c_{+}-
                                           b_{+}c_{+}b_{-}-
                                           c_{+}b_{+}b_{-}+
                                           2b_{+}b_{+}c_{-})$&$\frac{4}{3}\mu_{b}-\frac{1}{3}\mu_{c}$\\
$\Omega^{*0}_{bbc}$&$\frac{1}{\sqrt{3}}(b_{+}b_{+}c_{+}+b_{+}c_{+}b_{+}+c_{+}b_{+}b_{+})$&$2\mu_{b}+\mu_{c}$\\
$\Omega^{*-}_{bbb}$&$b_{+}b_{+}b_{+}$&$3 \mu_{b}$\\
\hline
\end{tabular}
 \ \ \ \ \ \ \ \ \ \ \ \ \ \ \ \ \ \ \ (* indicates $J^P=\frac{3}{2}^+$ state.)
\end{table}
\begin{table}[h]
 
\caption{Magnetic moments of triply heavy baryons in terms of
Nuclear magneton $\mu_{N}$} \label{tab:03}

\begin{tabular}{llllll}
\hline
&\multicolumn{3}{c}{\textbf{\underline{Potential index $p$}}}\\
Baryon &0.5&1.0&1.5&NRQM\cite{Amand2006}&NRQM \cite{Silvestre1996}\\
\hline
$\Omega^{++}_{ccc}$&1.149 &1.189 &1.190&$-$&1.023\\

$\Omega^{+}_{ccb}$&0.492 &0.502 &0.503&0.510&0.475\\
$\Omega^{*+}_{ccb}$   &0.637 &0.651 &0.652&$-$&$-$\\

$\Omega^{+}_{bbc}$&-0.199 &-0.203 &-0.203&-0.200&-0.193\\
$\Omega^{*+}_{bbc}$   &0.212 &0.216 &0.216&$-$&$-$\\

$\Omega^{-}_{bbb}$&-0.192 &-0.195 &-0.195&$-$&-0.180\\
\hline
\end{tabular}
\end{table}
If the interaction potential is hyper central symmetric such that
the potential depends on the hyper radius \,$x$\,\,only, then the
hyper radial schrodinger equation corresponds to the hamiltonian
given by Eqn(\ref{eq:404}) can be written as
\begin{equation}\label{eq:401}
\left[\frac{d^2}{dx^2}+\frac{5}{x}\frac{d}{dx}-\gamma(\gamma+4)\right]\phi_\gamma(x)=-2m[E-V(x)]\,\phi_\gamma(x)\\
\end{equation}\label{}
where $\gamma$ is the grand angular quantum number, m is the reduced
mass defined by
\begin{equation}\label{}
m=\frac{2\,\,m_\rho\,\, m_\lambda}{m_\rho+m_\lambda}
\end{equation}\label{}
For the present study we consider the hyper central potential $V(x)$
as\, \cite{Bhavin2008,BhavinPramana2008}
\begin{equation}\label{eq:410}
V(x)=-\frac{\tau}{x}+\beta x^p+\kappa+V_{spin}\left(x\right)\\
\end{equation}\label{}
In the above equation the first three terms correspond to
confinement potential in the hyperspherical co-ordinates. It belongs
to a general potential of the form $-Ar^{-\alpha}+kr^\epsilon+V_{0}
$ where $A,k,\alpha$ and $\epsilon$ are non negative constants where
as $V_{0}$ can have either sign. There are many potentials of this
generally, but with various values of the parameters have been
proposed for the study of hadrons \cite{Sameer2006}. For example,
Cornell potential has $\alpha=\epsilon=1$, Lichtenberg potential has
$\alpha=\epsilon=0.75$, Song-Lin potential has $\alpha=\epsilon=0.5$
and the Logarithmic potential of Quigg and Rosner corresponds to
$\alpha=\epsilon\rightarrow0$ etc have already been employed for the
study of hadron properties
\cite{Sameer2003,Sameer2004,Sameer0504107,Sameer0507209,Sameer1992,Sameer19921}.
 Martin potential
corresponds to  $\alpha=0,\epsilon=0.1$
\cite{Sameer2003,Sameer2004,Sameer0504107} while Grant-Rosner and
Rynes potential corresponds to  $\alpha=0.045$, $\epsilon=0$,
Heikkil\"{a}, T\"{o}rnqusit and Ono potential corresponds to
$\alpha=1$, $\epsilon=2/3$ \cite{Heikkila1984}. It is also been
explored in the region $0\leq \alpha \leq 1.2$, $0\leq \epsilon\leq
1.1$ of $\alpha-\epsilon$ values \cite{Song1991}. So, it is
important to study the behavior of the different potential scheme
with different choices of $\alpha$ and $\epsilon$ to know the
dependence of their parameters to the hadron properties. The
potential defined by Eqn(\ref{eq:410}) corresponds to $\alpha=1$ and
$\epsilon=p$. Here $\tau$ of hyper Coulomb, $\beta$ of confining
term and $\kappa$ are the model parameters.
  The parameter $\tau$ is related to the strong
running coupling constant $\alpha_{s}$ as
\cite{Bhavin2008,BhavinPramana2008}
\begin{equation}\label{}
\tau=\frac{2}{3}\,b\,\alpha_s
\end{equation}\label{}
where b is the model parameter and $\frac{2}{3}$ is the color factor
for the baryon and $\beta\approx m\tau$ numerically in terms of
(Me$V)^{p+1}$ has been used. The strong running coupling constant is
computed using the relation
\begin{equation} \label{}
\alpha_s=\frac{\alpha_s(\mu_0)}{1+\frac{33-2\,n_f}{12\,\pi}\alpha_s(\mu_0)ln(\frac{\mu}{\mu_0})}
\end{equation}\label{}
where $\alpha_s(\mu_{0}=1 GeV)\approx 0.6$ is considered in our
study. The forth term of Eqn(\ref{eq:410}) represents the spin
dependent part of the interaction. Unlike in our earlier study
\cite{BhavinPramana2008}, where the hypercentral spin-hyperfine
potential is parameterized without the explicit mass dependence of
the interacting quarks. Here, we consider the explicit mass
dependence as given by \cite{Bhavin2008,Garcilazo2007} and is taken
as
\begin{equation}\label{}
V_{spin}(x)=-\frac{1}{4}\alpha_s \, \frac{e^\frac{-x}{x_0}}{x^2
x_0}\sum\limits_{i{<}j}\frac{\vec{\sigma_i} \cdot
\vec{\sigma_j}}{6m_i m_j}\vec{\lambda_i}\cdot \vec{\lambda_j}
\end{equation}\label{}
The energy eigen value corresponding to Eqn(\ref{eq:401}) is
obtained using virial theorem for different choices of the potential
index $p$. The trial wave function is taken as the hyper coulomb
radial wave function given by \cite{Santopinto1998}
\begin{equation}\label{}
\psi_{\omega\gamma}=\left[\frac{(\omega-\gamma)!(2g)^6}{(2\omega+5)(\omega+\gamma+4)!}\right]^\frac{1}{2}(2gx)^\gamma
e^{-gx} L^{2\gamma+4}_{\omega-\gamma}(2gx)
\end{equation}\label{}
The baryon masses are determined by the sum of the quark masses plus
kinetic energy, potential energy and the spin dependent hyperfine
interaction as
\begin{equation}
M_B=\sum\limits_{i}m_i+\langle H \rangle + \kappa_{cm}
\end{equation}
Here $\kappa_{cm}$ corresponds to the center of mass correction. The
hyperfine interaction energy is treated here perturbatively.
 The computations are repeated for different choices of the flavour
combinations for $QQQ$ $(Q\ \epsilon \ b, c)$ systems. The computed
mass variations of these baryons without hyperfine interaction (spin
average mass) with respect to the potential index, $p$ are shown in
Fig \ref{fig:1}(a) , (b), (c) and (d) show the mass variations with
$p$ in the cases of $ccc$, $ccb$, $bbc$ and $bbb$ systems
respectively. Large variations in the computed masses are seen for
the choices of $p<1.0$, while in the cases of $p>1$, the mass found
to attain a saturated value in all the cases of the baryonic systems
studied here. Here we combine the potential parameter $\kappa$ with
the center of mass correction term $\kappa_{cm}$ as a single model
parameter, $E_0$ and is related to the total mass of the
constituting quarks. It is found that $E_0$ is linearly related to
the total mass of the system $(\sum\limits_{i}m_{i})$, as
\cite{Bhavin2008}
\begin{equation}
E_{0}=X \left(\sum\limits_{i}m_{i}\right) +Y
\end{equation}
where $X=$ 0.211 and $Y=$ -693.34  for all $QQQ$ combinations. The
mass parameter and hyperfine model parameter $x_{oqQQ}$ are taken
from our previous calculations \cite{Bhavin2008}.  We relate the
double heavy $x_{oqQQ}$ values to the triply heavy $x_{oQQQ}$ values
through an ansatz,
\begin{equation}
\frac{x_{o qQQ}}{x_{o
QQQ}}=A_{qQQ}\left(\frac{(\sum\limits_{i}m_{i})_{QQQ}}{(\sum\limits_{i}m_{i})_{qQQ}}\right)^{0.5}
\end{equation}
 with coefficient
$A_{qQQ}=0.5$ ($i.e, q=u,d$, $Q=c,b$ only)  for the non-strange
doubly heavy flavour baryons. Mass dependence on this parameter of
the hyperfine interaction has already been discussed in
\cite{Garcilazo2007} and \cite{Bhavin2008}. The computed masses of
the J$=\frac{1}{2}^+$ and J$=\frac{3}{2}^+$ states of the triple
heavy flavour baryons are listed in Table \ref{tab:01}in the range
of potential index  $0.5\leq p\leq2.0$ along with other theoretical
model
predictions.\\

\section{Effective quark mass and Magnetic moments of heavy baryons}
Generally, the meaning of the constituent quark mass corresponds to
the energy that the quarks carry inside the color singlet hadrons,
we call it as the effective masses of the quarks \cite{Close1982}.
Accordingly, the effective mass vary from system to system of
hadronic states. As the effective mass of the quarks would be
different from the adhoc choices of the model mass parameters. For
example, within the baryons the mass of the quarks may get modified
due to its binding interactions with other two quarks. Thus, the
effective mass of the $c$ and $b$ quark will be different when it is
in $ccb$ combinations or in $bbc$ combinations due to the residual
strong interaction effects of the bound systems.

Accordingly, we define
\begin{equation}\label{eq:417}
m^{eff}_{i}=m_i\left(
1+\frac{\left<H\right>+ \kappa_{cm}}{\sum\limits_{i}m_{i}}\right) \\
\end{equation}
Such that mass of the baryon,
\begin{equation}
M_B=\sum\limits_{i}m^{eff}_{i}\\
\end{equation}
Also, the magnetic moment of baryons are obtained in terms of its
constituent quarks as
\begin{equation}
\mu_B=\sum\limits_{i}\left<\phi_{sf}\mid\mu_{i}
\vec{\sigma}_{i}\mid\phi_{sf}\right>
\end{equation}
where
\begin{equation}
\mu_{i}=\frac{e_{i}}{2m_{i}^{eff}}
\end{equation}
Here $e_{i}$ and $\sigma_{i}$ represents the charge and the spin of
the quark$(\textbf{s}_{i}=\frac{\sigma_{i}}{2})$ constituting the
baryonic state and $\left|\phi_{sf}\right>$ represents the
spin-flavour wave function of the respective baryonic state
\cite{Simonov2002}. The details of the spin flavour combinations and
their wave functions corresponds to spin-$\frac{1}{2}$ and
spin-$\frac{3}{2}$ baryons are given in Table \ref{tab:02}. The
computed magnetic moments of the triply heavy flavour baryons are
listed in Table \ref{tab:03} for three choices of the potential
indices, 0.5, 1.0 and 1.5. Other existing model predictions are also
tabulated for comparision.

\section{Results and Discussion}
We have employed the hyper central model with hyperspherical
potential of the coulomb plus power potential form to study the
masses and magnetic moments of triply heavy flavour baryons$(QQQ,( Q
\ \epsilon \ c,\ b))$. It is important to see that the baryon mass
do not change appreciably beyond the potential power index $p
> 1.0$ (See Fig \ref{fig:1}(a) to (d)). For the present
calculation, we have employed the same mass parameters for the
$m_{c}$ and $m_{b}$ as used  the study of single heavy and double
heavy baryons \cite{Bhavin2008}. It is interesting to note that our
predictions of the mass of $QQQ$ baryons $( Q \ \epsilon \ c,\ b)$
are in good agreement with existing predictions based on other
theoretical models.\\
\\The predictions of the magnetic moment
of triply heavy flavour baryons studied here are with no additional
free parameters. Our results for magnetic moments of triply heavy
flavour baryons are listed in Table \ref{tab:03} and are compared
with other model predictions of \cite{Silvestre1996,Amand2006}. The
inter-quark interactions within the baryons are considered in the
calculation of magnetic moments through the definition of effective
mass of the constituent quarks within the baryon [Eqn
(\ref{eq:417})]. It is interesting to note that the magnetic moment
predicted in our model do not vary appreciably with different
choices of $p$ running from 0.5 to 1.5 as seen from Table
\ref{tab:03}. The predictions of $J=\frac{1}{2}$ baryons are also in
accordance with
 NRQM results \cite{Silvestre1996,Amand2006}.  \\
\\Experimental measurement of the heavy flavour baryonic magnetic
moments are sparse and few experimental groups (BTeV and SELEX
Collaborations) are expected to do measurements in near future.\\
\\We conclude that the three body description based on hyper central
co-ordinates and confinement potential assumed in this co-ordinate
has played a significant role bringing out a possible saturation
property of the basic interactions within the heavy baryons as seen
from the mass saturation of the baryons for the potential power
index $p > 1$. We also observed that  the model quark mass
parameters contributes significantly to the splitting of the
$J=\frac{3}{2}$ and $J=\frac{1}{2}$ states. The magnetic moments
predictions were found to be less sensitive to the potential power
index $p$. The disparity observed in our mass predictions with other
theoretical predictions can only be resolved with the experimental
confirmation of these states. We look forward to the experimental
support, from different heavy flavour production high luminosity experiments.\\
\\\textbf{Acknowledgement:} The authors acknowledge the financial support from University Grant commission,
Government of India under a Major research project F.
32-31/2006 (SR).\\

\end{document}